\documentclass[conference]{IEEEtran}

\usepackage{amsmath}
\usepackage{amssymb}
\usepackage{verbatim}
\usepackage{epsfig}

\long\def\comment#1{}

\newfont{\bb}{msbm10 scaled 1000}
\newcommand{\CC}{\mbox{\bb C}}

\newcommand{\EE}{\mbox{\bb E}}

\newfont{\bbsmall}{msbm10 scaled 700}

\newcommand{\hv}{{\bf h}}

\newcommand{\wv}{{\bf w}}

\newcommand{\xv}{{\bf x}}

\newcommand{\Hm}{{\bf H}}

\newcommand{\Cc}{{\cal C}}

\newcommand{\Nc}{{\cal N}}

\newcommand{\herm}{{\sf H}}

\def\argmax{\mathop{\rm argmax}}

\newcommand{\her}{{\sf H}}

\def\BibTeX{{\rm B\kern-.05em{\sc i\kern-.025em b}\kern-.08em
    T\kern-.1667em\lower.7ex\hbox{E}\kern-.125emX}}

\begin{document}

\title{Multi-User Diversity vs. Accurate Channel Feedback for MIMO Broadcast Channels}

\author{
\authorblockN{Niranjay Ravindran and Nihar Jindal}
\authorblockA{University of Minnesota \\
Minneapolis MN, 55455 USA\\
Email: \{ravi0022, nihar\}@umn.edu
}
} \maketitle \vspace{-1cm}

\begin{abstract}
A multiple transmit antenna, single receive antenna (per receiver) downlink
channel with limited channel feedback is considered. Given a constraint on
the total system-wide channel feedback, the following question is
considered: is it preferable to get low-rate feedback from a large number
of receivers or to receive high-rate/high-quality feedback from a smaller
number of (randomly selected) receivers? Acquiring feedback from many users
allows multi-user diversity to be exploited, while high-rate feedback
allows for very precise selection of beamforming directions. It is shown
that systems in which a limited number of users feedback high-rate channel
information significantly outperform low-rate/many user systems. While
capacity increases only double logarithmically with the number of users,
the marginal benefit of channel feedback is very significant up to the
point where the CSI is essentially perfect.
\end{abstract}

\section{Introduction}

Multiple antenna broadcast channels have been the subject of a tremendous
amount of research since the seminal work of Caire and Shamai
showed the sum-rate optimality of dirty-paper precoding (DPC) with Gaussian
inputs \cite{cs}.  If the transmitter is equipped with $M$ antennas, 
then multi-user MIMO techniques (such as DPC or sub-optimal but 
low-complexity linear precoding) that allow simultaneous
transmission to multiple users over the same time-frequency resource
can achieve a multiplexing gain of $M$ (as long as there are $M$ or
more receivers) even if each receiver has only one antenna.  
In contrast, orthogonal techniques (such as TDMA) that only serve 
one user achieve a multiplexing gain of only one.

Since the multiple antenna broadcast channel is a very natural
model for many-to-one communication (e.g., a single cell in a
cellular system), this line of work has been of great interest
to both academia and industry.  The
multiple antenna broadcast channel with \textit{limited channel feedback}
has been of particular interest over the past few years
because this accurately models the practical scenario where each
receiver feeds back (imperfect) channel information to the
transmitter.  In a frequency-division duplexed system (or a
time-division duplex system without accurate channel reciprocity)
channel feedback is generally the only mechanism by which the
transmitter can obtain channel state information (CSI). In
the single receive antenna setting, most proposed feedback strategies
either directly or indirectly involve each receiver quantizing 
its $M$-dimensional channel vector to the closest of a set of
quantization vectors; finer quantization corresponds to a larger
set of quantization vectors and thus higher rate channel feedback.

Within the literature on the MIMO broadcast with limited
feedback, there has been a dichotomy between the
extremes of systems with a small number of receivers
(on the order of the number of transmit antennas) versus systems
with an extremely large number of receivers.
\begin{itemize}
\item \textit{Finite systems} have been shown to be \textit{extremely}
sensitive to the accuracy of the CSIT, and thus require
\textit{high-rate feedback}.  This has been
shown from a fundamental information theoretic perspective 
\cite{lsw}, as well as in terms of particular
transmit strategies.  In particular, zero-forcing beamforming
has been shown to require CSIT quality that scales proportional
to SNR \cite{jindal}\cite{dl}.  

\item \textit{Large systems} have been shown to be able to operate 
near capacity with extremely \textit{low-rate channel feedback} in the 
asymptotic limit as the number of users is taken to infinity.  
In particular, \textit{random beamforming} (RBF) \cite{rbf} can operate with only 
$\log_2 M$ bits of feedback per user (plus one real number).
The performance of this technique in the asymptotic limit is
quite amazing: not only does the ratio of random beamforming
throughput to perfect CSIT capacity converge to one as 
the number of users is taken to infinity, but 
the difference between these quantities actually has been shown
to converge to zero \cite{sh}.

\end{itemize}

Finite systems require high-rate feedback because imperfect
CSIT leads to multi-user interference that cannot be resolved
at each receiver.  In order to prevent such a system from becoming
interference-limited, the CSIT must be very accurate; in terms of
channel quantization, this corresponds to using a very rich quantization
codebook that allows the direction of each receiver's channel
vector to be very accurately quantized.
In large systems, on the other hand, \textit{multi-user diversity}
is exploited to allow the system to operate with extremely low
levels of feedback.  The RBF strategy involves a quantization codebook
consisting of only $M$ orthonormal vectors (e.g.\@, the elementary basis vectors).  If such a codebook is used with
a small user population, each user's quantization will likely be quite
poor due to the limited size of the quantization codebook.
However, as the number of 
users increases, it becomes more and more likely that at least some
of the users have channel vectors that lie very close to one of
the $M$ quantization vectors.  This effect allows the system to
get by with very low rate feedback. Although the RBF throughput does
converge in the strong absolute sense to the perfect CSIT capacity,
convergence is extremely slow, even for systems with a 
small number of transmit antennas. 

Motivated by the apparent dichotomy between finite and asymptotically
large MIMO broadcast systems with limited channel feedback, in this
paper we ask the following simple question:\\
\textit{Is it preferable to have a system with a large number of receivers
and low-rate feedback from each receiver (thereby exploiting
multi-user diversity), or to have a system with a smaller number of receivers
with high-rate feedback from each receiver (thereby exploiting
the benefits of accurate CSIT)?} 

In order to fairly compare these systems, we equalize
the total number of channel feedback bits (across users).  Assuming that a total
of $T$ feedback bits are used, we compare the following:
\begin{itemize}
\item Random beamforming is used with $\frac{T}{\log_2 M}$ receivers
feeding back $\log_2 M$ bits each (in addition to one real number).
\item $\frac{T}{B}$ receivers quantize their channel direction to 
$B$ bits and feed back this information (plus one real number) to the transmitter, 
who uses a low-complexity user selection plus zero-forcing transmission strategy.
The parameter $B$ is varied within $\log_2 M \leq B \leq \frac{T}{M}$.
\end{itemize}
In performing this comparison, we assume the subset of users who
feedback are selected according to some \textit{channel-independent} 
criterion.  For example, they could be completely
randomly selected beforehand by the base station or the subset could
be chosen as the users with the largest user weights in a weighted
sum rate maximization setting.  

Our main conclusion is simple but striking: for almost
any number of antennas $M$ and SNR level, \textbf{system throughput is 
maximized by choosing $B$ (feedback bits per user) such that 
near-perfect CSIT is obtained for each of $\frac{T}{B}$ users that do feedback.}
For example, in a 4 antenna ($M=4$) system operating at 10 dB with
$T=100$ bits, the optimal is (approximately) achieved by having 4 users feedback 25
bits each, and the advantage relative to RBF (which involves 50 users feeding 
back $\log_2 M = 2$ bits each) is approximately 2.8 bps/Hz (9.6 vs. 6.8 bps/Hz). Note that $B = 25$ corresponds to CSIT at approximately 99.7\% accuracy, which is orders of magnitude more accurate than current wireless systems.
For larger values of $T$, the optimum is still achieved in
the neighborhood of $B=25$, i.e., a fraction of the user population
feed back very accurate CSI, and the significant performance advantage is
maintained even for very large values of $T$.
For relatively small values of $T$, the optimal $B$ is reduced because it is
still desirable to have at least $M$ users feedback, but high-rate quantization
from a small number of users is still desirable (e.g., for $T=40$ having
4 users feedback 10 bits gives a considerably larger throughput than RBF with 20 users).
Multi-user diversity provides a throughput gain that is only double-logarithmic in the 
number of users (who feedback CSI), while the marginal benefit of increased channel 
feedback is much larger up to the point where essentially near-perfect CSIT 
(relative to the system SNR) is achieved (e.g., 25 bits when $M=4$ and the system
is at 10 dB).

\section{Prior Work}

Previous work \cite{huan}\cite{swo}\cite{cioffi}\cite{khand1} has studied situations where
the individual receivers determine whether or not to feedback
on the basis of their current channel conditions (i.e., channel
norm and quantization error).  If each receiver makes channel-dependent
decisions then the base station does not \textit{a priori} know who
is going to feedback or how many users will feedback, which could
potentially complicate system design (possible solutions include
using random-access for feedback or somehow piggybacking the
variable feedback load onto uplink data packets).  From only a throughput
maximization perspective, one would intuitively think that making channel-dependent
feedback decisions would perform better than channel-independent decisions, because
only users with strong channels and good quantization feed back. However, there
are other scenarios where channel-independent selection of users would be
preferable, e.g., when users have delay-sensitive traffic and are requested
to feed back when their deadlines are approaching.  There are many important
differences between the approaches and both have their strengths and
weaknesses. In this work, we consider only channel-independent approaches, although we expect to compare against channel-dependent approaches in the future.

Another recent work has studied the tradeoff between multi-user diversity
and accurate channel feedback in the context of two-stage feedback
\cite{gesbert}. In the first stage, all users feed back coarse estimates of
their channel, based on which the transmitter runs a selection algorithm to
select $M$ users who feedback more accurate channel quantization during the
second feedback stage. Our work differs in that we consider only a single
stage approach, and more importantly in that we optimize the number of
users ($T/B$ randomly selected users) who feed back accurate
information rather than limiting this number to $M$. Indeed, this
optimization is precisely why our approach shows such large gains over
naive RBF or un-optimized zero forcing.

\section{System Model \& Background}

We consider a multi-input multi-output (MIMO) Gaussian broadcast channel in which the Base Station (BS) or transmitter
has $M$ antennas and each of the $K$ users have 1 antenna each. The channel output $y_k$ at user $k$ is given by:
\begin{equation} \label{model}
y_k = \hv_k^\her \xv + z_k, \;\; k = 1,\ldots,K
\end{equation}
where $z_k \sim \Cc\Nc(0,1)$ models Additive White Gaussian Noise (AWGN), $\hv_k \in \CC^M$ is the vector of channel coefficients from the $k^\textrm{th}$ user antenna to the transmitter antenna array and $\xv$ is the vector of channel input symbols transmitted by the base station. The channel input is subject to the average power constraint $\EE[|\xv|^2] \leq P$.

We assume that the channel {\em state}, given by the collection of
all channel vectors $\Hm = [\hv_1,\ldots,\hv_K] \in \CC^{M \times
K}$, varies in time according to a block-fading model, where $\Hm$
is constant over each {\em frame}, and evolves from frame to frame according to an ergodic stationary spatially
white jointly Gaussian process, where the entries of $\Hm$ are Gaussian i.i.d. with elements
$\sim \Cc\Nc(0,1)$.

Each user is assumed to know its own channel perfectly. At the beginning of each block, each user quantizes its channel 
to $B$ bits and feeds back the bits perfectly and instantaneously to the 
access point. Vector quantization is performed using a codebook ${\mathcal C}$ 
that consists of $2^{B}$ $M$-dimensional unit norm vectors
${\mathcal C} \triangleq \{ \mathbf{w}_1, \ldots, \mathbf{w}_{2^{B}} \}$.
Each user quantizes its channel vector to the quantization vector that forms the minimum angle to it.
Thus, user $k$ quantizes its channel to $\widehat{\hv}_k$, chosen according to:
\begin{eqnarray} \label{eq-quant}
\widehat{\hv}_k & = & \textrm{arg} \min_{\wv \in {\mathcal C}}\ 
\sin^2 \left( \angle ({\bf h}_k, \wv) \right).
\end{eqnarray}
and feeds the quantization index back to the transmitter. In addition to this, each user also feeds back a single real number, which can be the channel norm, or some other channel quality indicator.

We assume that a total of $T$ bits are allocated for feedback, and that there are at least $\frac{T}{\log_2 M}$ users available to feedback CSI, if needed. The following feedback strategies are considered:

\subsection{Random Beamforming}

The Random beamforming scheme proposed in \cite{rbf} is used, where each user feeds back $\log_2 M$ bits along with one real number. The number of users feeding back information is hence $\frac{T}{\log_2 M}$. In this case, ${\mathcal C}$ consists of $M$ orthogonal unit vectors, and the codebook is common to all users. In addition to the quantization index, each user feeds back a real number representing its SINR, should it be selected. If $\wv_m$ ($1 \leq m \leq 2^B = M$) is selected to be the `best' quantization vector for user $k$, where $1 \leq k \leq \frac{T}{\log_2 M}$, the SINR for the user is:
\begin{eqnarray} \label{RBF-SINR}
\text{SINR}_{k, m} & = & \frac{|\hv_k^\herm\wv_m|^2}{\frac{M}{P} + \sum\limits_{n\neq m}|\hv_k^\herm\wv_n|^2}.
\end{eqnarray}
`Simple' user selection is used, i.e.\@, the user with the highest SINR on each $\wv_m$ is chosen, and $\wv_1, \dots, \wv_M$ are used as the beamformers. This constitutes a simple and low-complexity user-selection algorithm.

\subsection{Random Vector Quantization}

We consider the case when $\frac{T}{B}$ users quantize their channel direction to 
$B$ bits and feed back this information to the transmitter, along with the channel norm $||\hv_k||^2$. Here, ${\mathcal C}$ consists of random unit-vectors independently chosen from the isotropic distribution on the $M$-dimensional unit sphere \cite{honig} (RVQ). Each user is assumed to use a different and independently generated codebook\footnote{Note that random vector quantization allows us to simulate large quantization codebooks using the statistics of the quantization error (which is known), permitting a Monte Carlo simulation}.
The transmitter uses low-complexity greedy user selection \cite{dimic} along with zero-forcing transmission, where the quantized channel (i.e.\@, the channel $||\hv_k||\cdot\widehat{\hv}_k$) is treated as if it were the true channel, for user selection purposes. We consider only the case when the channel norm information $||\hv_k||^2$ is fed back, as opposed to (the receiver's estimate of) the SINR, which may take quantization error into account \cite{yoo}.

The parameter $B$ is varied within $1 + \log_2 M \leq B \leq \frac{T}{M}$. In general, if $R_\textsc{ZF-RVQ}(P, M, K, B)$ represents the ZF rate for a system with $M$ antennas at the transmitter, SNR $P$ and $K$ users, each feeding back $B$ bits (in addition to one real number), the optimal $B$ is found as follows:
\begin{equation}
B^\textsc{OPT} = \argmax_{1+\log_2M \leq B \leq \frac{T}{M}}\ R_\textsc{ZF-RVQ}\left(P, M, \frac{T}{B}, B\right)
\end{equation}

\vspace{11pt}

Random beamforming involves the maximum number of users $\left(\frac{T}{\log_2 M}\right)$
but the minimum number of feedback bits per user ($\log_2 M$), while
the ZF strategy can vary from a large system with low-rate feedback 
($B=1 + \log_2 M$) all the way to a small system with very high-rate feedback
($M$ users, $B = T/M$).

\section{Basic Results and Intuition}

To gain an understanding of the optimal $B$, we propose the following approximate characterization. We model the rate expression in terms of the parameters $P, M, B$ and $T$ as follows:
\begin{eqnarray}
R_\textsc{ZF-Approx}\left(P, M, \frac{T}{B}, B\right) = M \log_2\left(\frac{P}{M}\log_2\left(\frac{T}{B}\right)\right) \nonumber\\ -M \log_2\left(1 + \frac{P}{M}\log_2\left(\frac{T}{B}\right) 2^{-\frac{B}{M-1}}\right)
\end{eqnarray}

\begin{figure}[ht]
\begin{center}
\includegraphics[width = 3.7in]{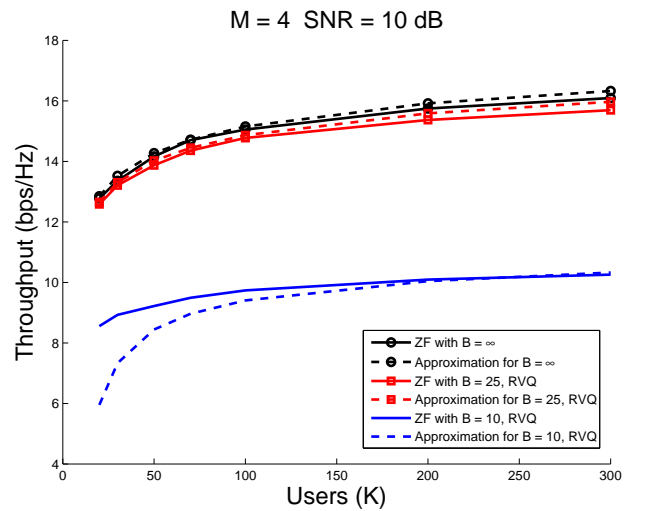}
\caption{Accuracy of the approximate throughput expression}
\label{fig7}
\end{center}
\end{figure}

The $M\log_2\left(\frac{P}{M}\log_2\left(\frac{T}{B}\right)\right)$ term captures the effect of multiuser diversity due to $\frac{T}{B}$ users (as well as appropriate scaling with SNR and $M$) for ZF with perfect CSIT. This is asymptotically correct, to an $O(1)$ term \cite{yoo2}. The $M \log_2\left(1 + \frac{P}{M}\log\left(\frac{T}{B}\right) 2^{-\frac{B}{M-1}}\right)$ term serves to capture the throughput loss due to limited channel feedback, relative to perfect CSIT. The effect of finite rate feedback was quantified to be $\EE\left[M\log_2\left(1 + \frac{P}{M}|\hv_k||^2 2^{-\frac{B}{M-1}}\right)\right]$ in \cite{jindal}, for a $K \leq M$ user system (i.e., without user selection). This is applied for a $K = \frac{T}{B} > M$ user system by noting that the quantization error remains unaffected in spite of $K > M$ users (as quantization error information is not fed back). However, we note that due to user selection, $\frac{P}{M} ||\hv_k||^2$ behaves as $\frac{P}{M}\log_2\left(\frac{T}{B}\right)$ when $\frac{T}{B}$ users are involved. This also captures the fact that keeping $B$ fixed and taking $T$ to $\infty$ (for a fixed $P$) will essentially nullify all multiuser diversity making the system interference limited, as described in \cite{yoo}.  Figure \ref{fig7} depicts the accuracy of the approximate expression for an $M = 4$ system at $10$ dB. Note that there may still be an $O(1)$ constant error, but this is irrelevant for our optimization. 

Based on this expression, an approximate expression for $B^\textsc{OPT}$ may be computed as:
\begin{eqnarray} \label{opt-approx}
\widehat{B}^\textsc{OPT} = & \argmax\limits_{B} & \log_2\left(\log_2\left(\frac{T}{B}\right)\right) -\nonumber\\ & & \log_2\left(1 + \frac{P}{M} \log_2\left(\frac{T}{B}\right) 2^{-\frac{B}{M-1}}\right)
\end{eqnarray}

The solution to this problem is obtained by solving:
\begin{eqnarray} \label{minexpr}
\frac{M-1}{M} P 2^{-\frac{\widehat{B}^\textsc{OPT}}{M-1}} \widehat{B}^\textsc{OPT} \left(\log_e\left(\frac{T}{\widehat{B}^\textsc{OPT}}\right)\right)^2 = 1
\end{eqnarray}
This expression is obtained by equating the derivative of (\ref{opt-approx}) to zero, and solving for $B$.

\begin{figure}[ht]
\begin{center}
\includegraphics[width = 3.7in]{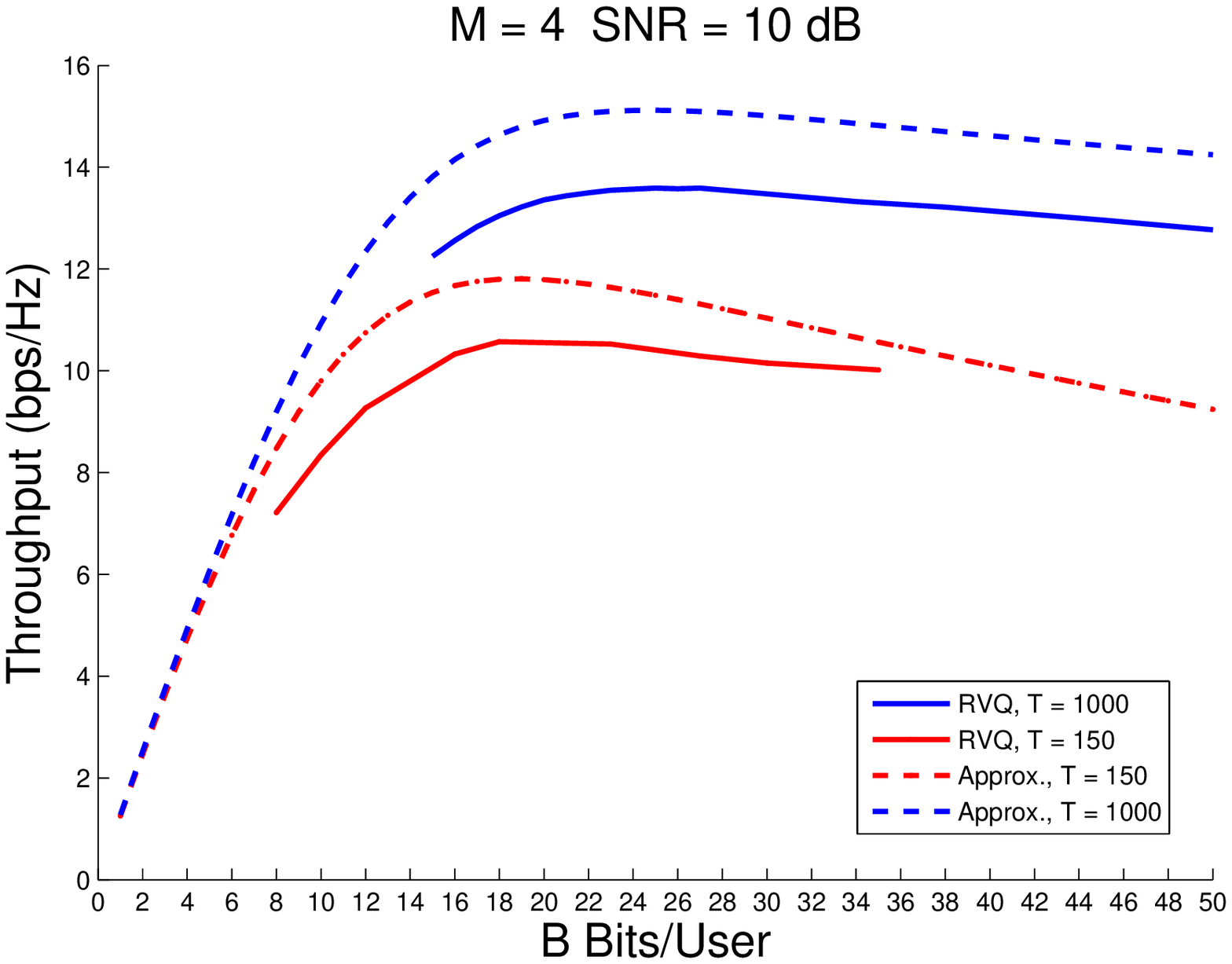}
\caption{Behavior of $B^\textsc{OPT}$ with $T$}
\label{fig6}
\end{center}
\end{figure}

In Figure~\ref{fig6}, the true throughput $R_\textsc{ZF-RVQ}\left(P, M, \frac{T}{B}, B\right)$ and the approximation $R_\textsc{ZF-Approx}\left(P, M, \frac{T}{B}, B\right)$ are plotted (versus $B$) for an $M = 4$ system at $10$ dB SNR with $T = 150, 1000$ bits. For $T = 150$, $B^\textsc{OPT} = 18, \widehat{B}^\textsc{OPT} = 19$ and for $T = 1000$, $B^\textsc{OPT} = \widehat{B}^\textsc{OPT} = 25$. In both cases, the approximation yields relatively accurate results. Also note that the throughput grows rapidly for smaller values of $B$, but falls relatively slowly after the optimal $B$ has been attained, and there is not much difference in performance in this region.

\begin{figure}[ht]
\begin{center}
\includegraphics[width = 3.7in]{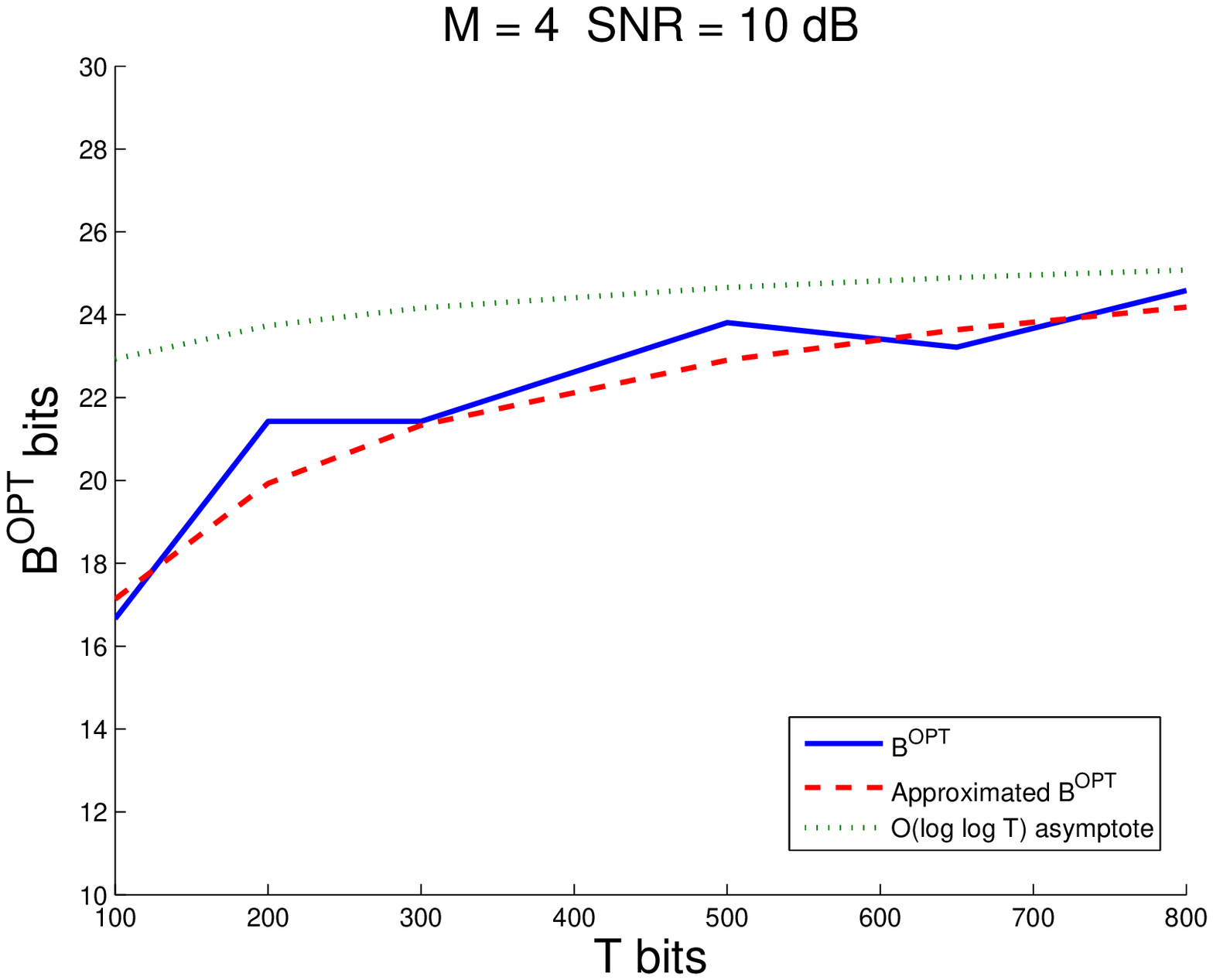}
\caption{Behavior of $B^\textsc{OPT}$ with $T$}
\label{fig1}
\end{center}
\end{figure}

Figure~\ref{fig1} depicts the behavior of $B^\textsc{OPT}$ with $T$. $\widehat{B}^\textsc{OPT}$ is seen to reasonably capture the behavior of $B^\textsc{OPT}$, and this dependence is numerically found to be $B^\textsc{OPT} \sim O(\log(\log(T)))$. This intuitively makes sense, as this would mean that $2^{-\frac{B}{M-1}} \sim O(1/\log(T))$ which would compensate for the $\log_2\left(\frac{T}{B}\right)$ term in the interference portion of (\ref{opt-approx})\footnote{It was observed in \cite{yoo} that pure `norm' information used for user selection (i.e.\@, without taking the quantization error magnitude into account) would cause the system to become interference limited (as the number of users feeding back are taken to infinity). However, selection of an optimal $B$ may be able to overcome this disadvantage.}. Furthermore, this growth rate also implies that $B^\textsc{OPT}$ grows extremely slowly for larger values of $T$, and one would prefer essentially the same feedback quality even if $T$ is very large.

\begin{figure}[ht]
\begin{center}
\includegraphics[width = 3.7in]{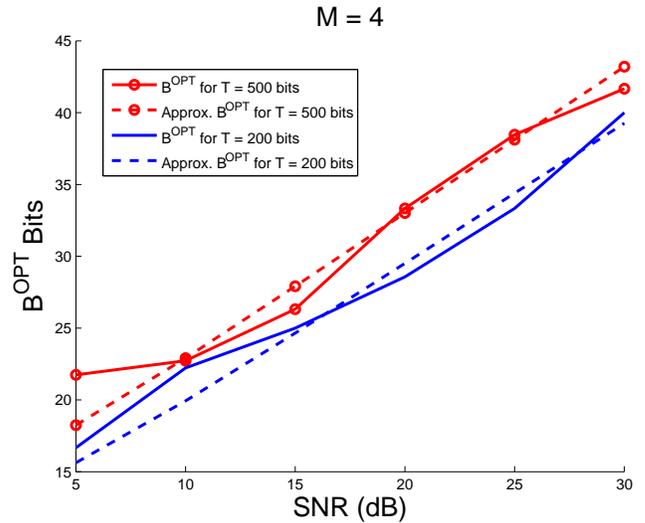}
\caption{Behavior of $B^\textsc{OPT}$ with SNR $P$}
\label{fig2}
\end{center}
\end{figure}

It is similarly observed that $B^\textsc{OPT}$ scales linearly with the system SNR and $M$, i.e.\@, $B^\textsc{OPT} \sim O(M \log(P))$, which is seen in Figure \ref{fig2}. The approximate expression $\widehat{B}^\textsc{OPT}$ is seen to accurately model this behavior as well. Interestingly, this behavior of the number of feedback bits is the same as with an $M$-user system \cite{jindal} (without user selection). Further, this also suggests that a smaller fraction of users should feedback as SNR grows, and at large SNR there would essentially be only $M$ users feeding back with $\frac{T}{M}$ bits each.







\section{Simulation Results}

\begin{figure}[ht]
\begin{center}
\includegraphics[width = 3.7in]{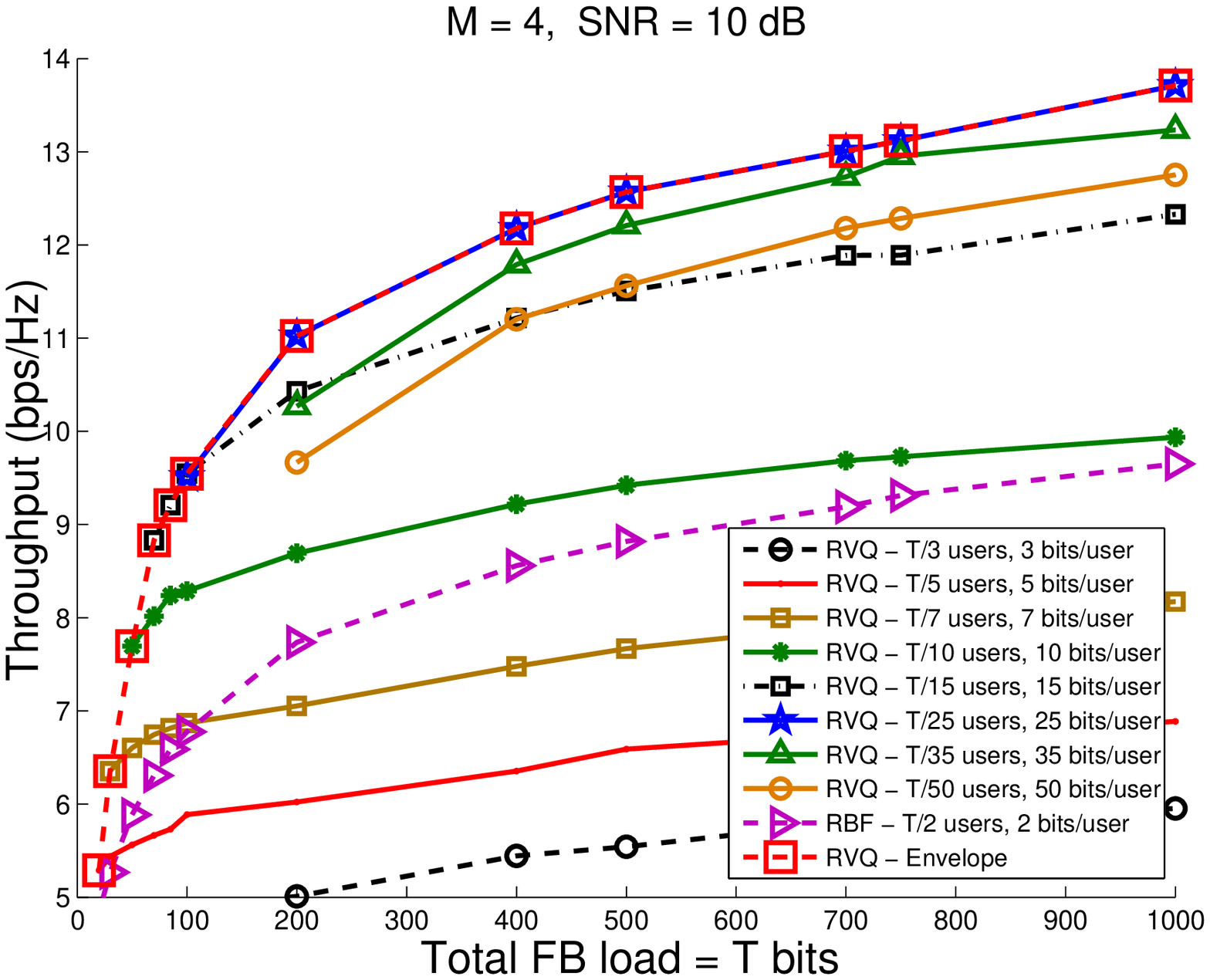}
\caption{RBF vs. Optimized number of feedback users, $M = 4$}
\label{fig3}
\end{center}
\end{figure}

In a 4 antenna ($M=4$ system, Figure \ref{fig3}), $R_\textsc{ZF-RVQ}\left(P, M, \frac{T}{B}, B\right)$ is plotted versus $T$ for various values of $B$. For each choice of $B$, $\frac{T}{B}$ users feed back information. Random vector quantization with zero forcing and greedy selection are used, as described previously. This is compared with Random beamforming with a fixed codebook size of $2$ bits. At an SNR of $10$ dB with a total budget of $T = 100$ bits for feedback, the optimal is (approximately) achieved when $4$ users each feedback $25$ bits worth of information. 

For larger values of $T$, the optimum is still (approximately) achieved in
the neighborhood of $B^\textsc{OPT} = 25$, i.e., a fraction of the user population
feed back very accurate CSI. It is seen that there is a significant performance advantage relative to RBF. This advantage is expected to diminish as $T$ grows, but it is seen that the significant advantage is maintained even for very large values of $T$ ($5000$ bits and above). The value of $B^\textsc{OPT}$ grows very slowly beyond 25 as $T$ increases, which agrees with the $O(\log(\log(T)))$ expression.

\begin{figure}[ht]
\begin{center}
\includegraphics[width = 3.7in]{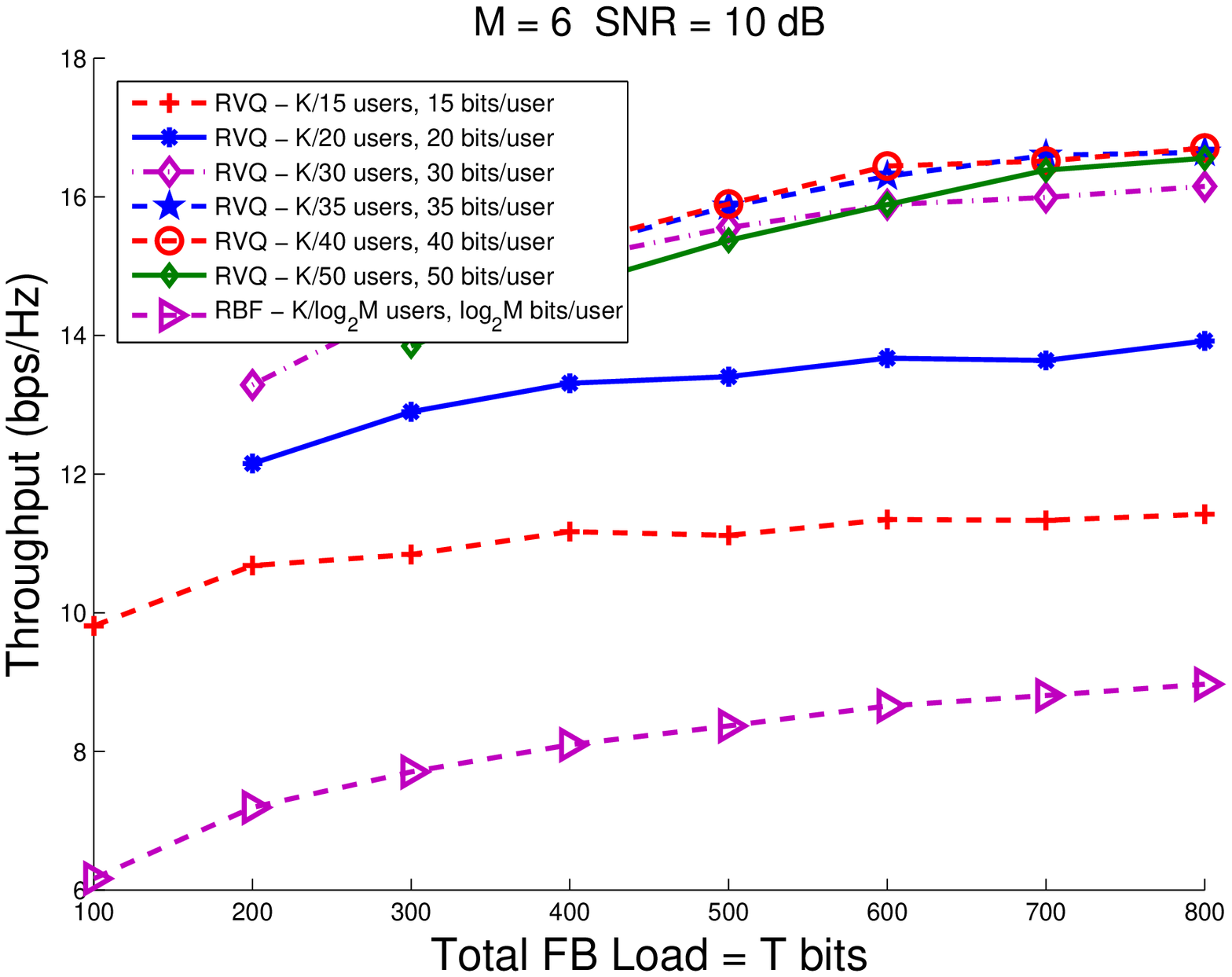}
\caption{RBF vs. Optimized number of feedback users, $M = 6$}
\label{fig4}
\end{center}
\end{figure}

Similar behavior is observed in an $M = 6$ system in Figure~\ref{fig4}. The optimal number of bits is approximately $35$ (as opposed to 25 for $M = 4$) for larger values of $T$.

\begin{figure}[ht]
\begin{center}
\includegraphics[width = 3.7in]{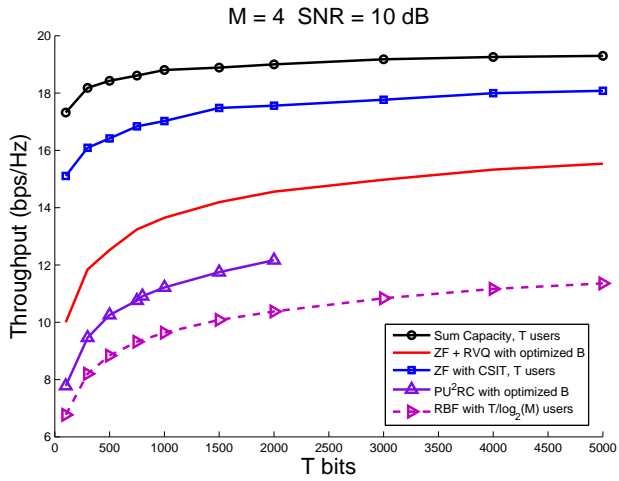}
\caption{RBF vs. Optimized number of feedback users, $M = 4$, Large T}
\label{fig5}
\end{center}
\end{figure}

Figure~\ref{fig5} depicts the performance of the random vector quantization scheme with optimized $B$, for very large $T$. This is compared to the sum capacity of the $T$-user system with CSIT (computed using the iterative waterfilling algorithm \cite{iwf}) as well as Zero forcing with greedy selection among $T$ users and perfect CSIT. The advantage relative to random beamforming is maintained, due to the slow convergence of RBF. As a generalization of random beamforming, $\text{PU}^2\text{RC}$ beamforming is also considered. This scheme uses several sets of codebooks, each codebook consisting of $M$ orthogonal unit-vectors (the RBF codebook). If $2^{B - \log_2 M}$ such sets are used, the total number of bits per user is $B$. Just as in RBF, all users have the same set of codebooks and each user feeds back the index of its `best' quantization as well as the SINR. The transmitter performs the same simple selection as random beamforming for each of the $2^{B - \log_2 M}$ codebooks, and then picks the one that maximizes the rate. Just as with the random vector quantization scheme, $B$ is optimized so that $\frac{T}{B}$ users feedback $B$ bits each. While this scheme should perform strictly better than random beamforming, there is still a significant gap relative to random vector quantization and zero forcing with optimized $B$.

\section{Conclusion}  
In this paper we have considered the very simple but apparently overlooked
question of whether low-rate feedback/many user systems or high-rate
feedback/limited user systems are preferable in the context of MIMO
downlink channels. Answering this question essentially reduces to comparing
the value of multi-user diversity (many users) versus channel information
(high-rate feedback), and the surprising conclusion reached is that there
is an extremely strong preference towards accurate channel information.
Although there may be other issues that influence the design of channel
feedback protocols, this work suggests that very high-rate channel feedback
should receive serious consideration if multi-user MIMO techniques are
employed on the downlink channel.

\bibliographystyle{IEEEtran}

\end{document}